\documentclass[10pt, conference, letterpaper]{IEEEtran}

\pdfoutput=1
\usepackage{algorithmic, algorithm}
\usepackage[cmex10]{amsmath}
\usepackage{amssymb, latexsym}
\usepackage[caption=false,font=footnotesize]{subfig}
\usepackage{cite}
\usepackage{pgfplots}

\pgfplotsset{compat=newest}
\pgfplotsset{plot coordinates/math parser=false}
\newlength\figureheight
\newlength\figurewidth

\begin{document}

\title{NetCodCCN: a Network Coding approach for Content-Centric Networks}

\author{ \IEEEauthorblockN{Jonnahtan Saltarin\IEEEauthorrefmark{1}, Eirina Bourtsoulatze\IEEEauthorrefmark{1}, Nikolaos Thomos\IEEEauthorrefmark{2} and Torsten Braun\IEEEauthorrefmark{1}}
\IEEEauthorblockA{\IEEEauthorrefmark{1}University of Bern, Bern, Switzerland\\ Email: \{saltarin,braun\}@inf.unibe.ch, eirina.bourtsoulatze@gmail.com}
\IEEEauthorblockA{\IEEEauthorrefmark{2}University of Essex, Colchester, United Kingdom\\ Email: nthomos@essex.ac.uk}}

\maketitle

\begin{abstract}
Content-Centric Networking (CCN) naturally supports multi-path communication, as it allows the simultaneous use of multiple interfaces (e.g. LTE and WiFi). When multiple sources and multiple clients are considered, the optimal set of distribution trees should be determined in order to optimally use all the available interfaces. This is not a trivial task, as it is a computationally intense procedure that should be done centrally. The need for central coordination can be removed by employing network coding, which also offers improved resiliency to errors and large throughput gains. In this paper, we propose NetCodCCN, a protocol for integrating network coding in CCN. In comparison to previous works proposing to enable network coding in CCN, NetCodCCN permit Interest aggregation and Interest pipelining, which reduce the data retrieval times. The experimental evaluation shows that the proposed protocol leads to significant improvements in terms of content retrieval delay compared to the original CCN. Our results demonstrate that the use of network coding adds robustness to losses and permits to exploit more efficiently the available network resources. The performance gains are verified for content retrieval in various network scenarios.
\end{abstract}

\section{Introduction}
\label{sec:introduction}
In the IP protocol, core of the current Internet architecture, each packet is routed based on the location of the host to which it is addressed. However, nowadays Internet users care more about the content they want to obtain rather than where this content is stored. To address this mismatch, Jacobson \emph{et al.}~\cite{Jacobson2009} proposed Content-Centric Networking (CCN), a new communication paradigm in which the importance is shifted from \textit{where} the content is located, to \textit{what} the content is. In the CCN model, the content is described by its \textit{name} and the users demand content with the help of Interest messages that contain the name of the requested content. These Interests are transmitted over the network until they reach a node holding a copy of the content whose name matches that of the Interest message. Once such a node is reached, a copy of the requested content is encapsulated in a Data message and it is sent back to the requester following the reverse path of that followed by the Interest message. As the Data message is transmitted backwards to the requester, intermediate nodes can store copies of it, so they can reply to future Interests for the same content.

One of the advantages of CCN is that it allows clients to exploit multiple paths in a native way, as clients can simultaneously transmit Interests over all their network interfaces (\textit{e.g.}, LTE and WiFi) to retrieve the content segments that comprise the requested content. This leads to a better use of the network resources and reduces the time needed to collect all the content segments. However, when multiple clients are interested in the same content (\textit{e.g.}, a popular video stream), and/or when the content is distributed across multiple sources (\textit{e.g.}, in a distributed storage system), the optimal content delivery rate is only attained if the segments are delivered over the optimal set of multicast trees~\cite{Wu2004}. This means that the Data and Interest messages should be transmitted over these multicast trees, thus, the nodes need to know where they need to forward each Interest to follow these multicast trees, which does not scale for large and dynamic topologies. Furthermore, the computation of the optimal set of multicast trees needs a central entity that is aware of the network topology, which is hard to be done in dynamic networks. An alternative solution to the computation of the optimal multicast trees is to use network coding~\cite{Ahlswede2000}. With network coding all the network nodes perform coding operations on the received packets instead of just replicating and forwarding them as in traditional networks. The receivers decode the information when they receive a decodable set of packets, \emph{i.e.}, as many linearly independent coded packets as the number of source data segments.


The application of network coding in CCN has been explored in~\cite{Montpetit2012} where the NC3N architecture has been introduced. In this approach, Interests contain information about the content segments available at the client, based on the approach proposed in~\cite{Sundararajan2011}. Nodes holding content segments that matches the name prefix of the Interest, reply only if they can provide a network coded content segment that increases the content available at the client. However, in the presence of multiple clients, \textit{(i)} the aggregation of Interests is problematic, since Interests for the same content segment from different clients contain different content availability information; and \textit{(ii)}, when a client sends multiple Interests in parallel to receive multiple content segments, it will include the same information about the content that it already has. This is undesirable as a node that has a matching content segment will reply to multiple Interests with the same content segment, that will be duplicated for the client. Inspired by~\cite{Montpetit2012}, CodingCache has been proposed in~\cite{Wu2013} which uses network coding to replace the content segments in the cache of the network nodes. Due to the increased content segment diversity in the network, the cache hit rate is improved. However, this approach suffers from the same drawbacks as the architecture in~\cite{Montpetit2012}. In \cite{Llorca2013}, the multicast delivery of network coded content in Information-Centric Networks is optimized by finding the evolution of the content segments that are stored in the network. The drawback of this approach is that it does not scale well with the number of network nodes, because it needs a central entity that is aware of the network topology and the clients' requests.

In this paper we propose NetCodCCN, a protocol for integrating network coding in CCN. Our proposed solution solves the shortcomings of the approaches presented in~\cite{Montpetit2012} and~\cite{Wu2013}. Specifically, \textit{(i)} we eliminate the need to include in the Interests the information about the content available at the client, thus, simplifying Interest aggregation; \textit{(ii)} we allow nodes to keep information about the content segments they have sent on each face, reducing the number of duplicate segments; and \textit{(iii)} we allow clients to send multiple Interests in parallel, by modifying the way in which the nodes process the Interest messages.


We have implemented NetCodCCN by making the necessary changes to CCNx~\cite{Ccnx2015}, and performed experiments to compare it to unmodified CCNx. Our results demonstrate that NetCodCCN offers large gains in terms of the time needed to retrieve the original content object. Moreover, it adds robustness to losses and permits to exploit more efficiently the available network resources in multi-source multicast scenarios. To the best of our knowledge, this is the first practical implementation of network coding integrated into CCNx.

\section{Data retrieval in CCN}
\label{sec:ccn}


We focus on content communication over wired networks represented by directed acyclic graphs $\mathcal{G = (V,E)}$, where $\mathcal{V}$ and $\mathcal{E}$ denote the set of network nodes and the set of links connecting them, respectively. Each network consists of a set of source nodes $\mathcal{S}$ that generate and/or store content objects, a set of clients $\mathcal{U}$ that demand content objects and a set of intermediate nodes $\mathcal{R}$ through which the content objects are requested and transmitted. Hence, we have $\mathcal{V = S \cup U \cup R}$, where every node $v \in \mathcal{V}$ is connected with its neighboring nodes through a set of faces $\mathcal{F}_{v}$.

In CCN, content objects are split into smaller segments that fit into Data messages. Each segment is uniquely identified by a name. We denote a content object as $C_{p} =\{ c_{p,1},\dots, c_{p,N} \}$ where $N$ is the number of segments in $C_{p}$ and $p$ is the name of the content object, which serves as a name prefix for the segments. The name of each segment $c_{p,n} \in C_{p}$ is generated by appending the name of the content object $p$ with the segment identifier $n$. For instance, the name of the segment $c_{p,1}$ is \texttt{/provider/videos/largevideo.h264/1}, where \texttt{/provider/videos/largevideo.h264} is the name prefix, $p$ and \texttt{1} is the segment id.

Each source $s \in \mathcal{S}$ stores content objects that can be requested by the clients. A client $u \in \mathcal{U}$ that is interested in a content object $C_{p} =\{ c_{p,1},\dots, c_{p,N} \}$ should send a set of Interest messages $I_{p} =\{ i_{p,1},\dots, i_{p,N} \}$, one for each segment. These interests are sent over a set of faces $\mathcal{F}_{u}^{p}$ that are configured to forward Interests for content with name prefix $p$. The information about which faces a node can use to send Interests for specific name prefixes is stored in the \emph{Forwarding Information Base (FIB)} table.

In CCN, each node $v \in \mathcal{V}$ has a cache, or \emph{Content Store (CS)} in CCN terminology, where segments that pass through the node can are stored. These segments can be used later to reply to Interests for segments with a matching name. Therefore, a node $v \in \mathcal{I}\cup\mathcal{S}$ holding a copy of the segment $c_{p,n}$ in its CS will reply to any Interest $i_{p,n}$. If the CS of node $v$ does not contain a segment matching the name of the Interest $i_{p,n}$, the node $v$ first checks its \emph{Pending Interest Table (PIT)}, that keeps track of the Interests forwarded by the node and all the faces over which those Interests have arrived. If the node $v$ finds in its PIT an entry that matches the name in the Interest, it knows that it has already forwarded $i_{p,n}$ and hence the segment $c_{p,n}$ is expected. In this case $v$ does not forward $i_{p,n}$ again, but only adds the face $f$ over which the Interest has arrived to the respective PIT entry. When the PIT does not have any entry that matches the Interest $i_{p,n}$, the node $v$ forwards the Interest to its neighboring nodes over the set of faces $\mathcal{F}_{u}^{p}$ configured in its FIB.

Once the requested segment is found in the CS of an intermediate node or in a source node, it is transmitted to the client in a Data message over the reverse path of that followed by the Interest. When a node $v$ receives a Data message with the segment $c_{p,n}$ over a face $f \in \mathcal{F}_{v}$, it first checks its CS. If a segment with the same name exists, the arrived segment $c_{p,n}$ is considered duplicated and is not transmitted further. If there is no matching segment in the CS, the node checks the PIT for an entry that matches the name of the segment $c_{p,n}$. If there is no matching PIT entry, the segment $c_{p,n}$ is considered unsolicited and it is discarded. If a matching PIT entry is found, the segment is forwarded to all the faces specified in the corresponding PIT entry. Additionally, the segment $c_{p,n}$ may be added to the CS, according to the caching policy.

\section{Towards Network coding enabled CCN}
\label{sec:towards_nc}

In this section we describe the benefits that network coding can bring to CCN. First, we motivate the use of network coding by presenting three scenarios in which CCN does not perform efficiently. Then, we show how network coding can alleviate the drawbacks of CCN in the mentioned scenarios, while also bringing additional benefits.

\subsection{Motivation}
\label{sec:motivation}

\begin{figure*}[t]
	\centering
	\subfloat[][]{\label{fig:example_1}\includegraphics[width=0.33\textwidth,page=1]{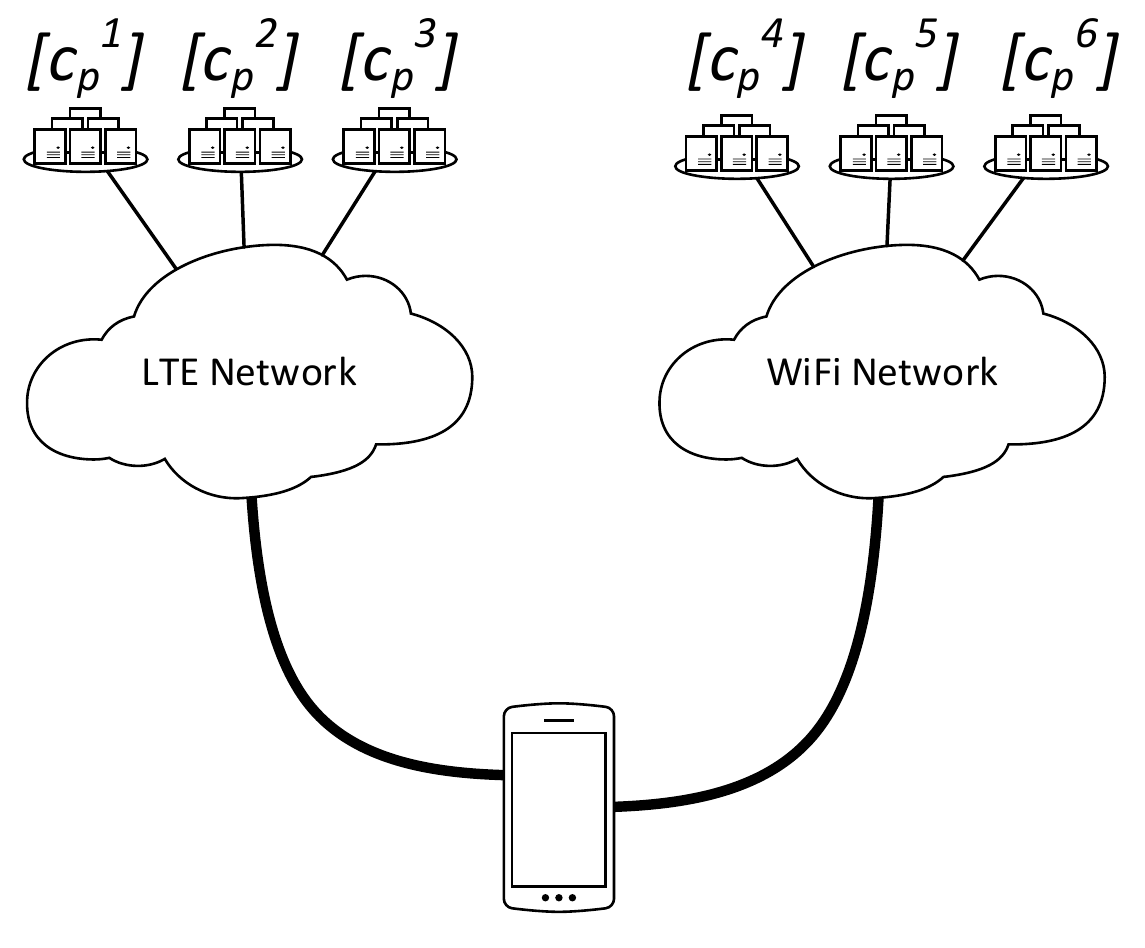}}
	\hspace{10pt}
	\subfloat[][]{\label{fig:example_2}\includegraphics[width=0.3\textwidth,page=2]{example.pdf}}
	\hspace{10pt}
	\subfloat[][]{\label{fig:example_3}\includegraphics[width=0.3\textwidth,page=3]{example.pdf}}
	\caption[]{Mobile devices retrieving segments in parallel over LTE and WiF): \subref{fig:example_1} multi-source unicast; \subref{fig:example_2} single-source multicast; \subref{fig:example_3} multi-source multicast (butterfly network).}
	\label{fig:example}
\end{figure*}

Nowadays, communication devices usually come with multiple network interfaces that can be used to gather content, \textit{e.g.}, smart-phones usually have WiFi and 3G/LTE interfaces. However, in the traditional host-centric networking, using multiple interfaces in parallel to retrieve content is a difficult task, as end-to end connections need to be established for each interface. In CCN, multipath content retrieval is naturally supported, as the clients can distribute all the Interest messages needed to retrieve a content object over all its available faces. However, there are some scenarios where CCN does not provide efficient support for multipath content retrieval:

\begin{itemize}

\item \textit{Multi-source unicast:} Let us consider the case illustrated in Fig.~\ref{fig:example_1}, where a client $u$ is interested in a content object $C_{p}$ composed of $N$ segments that are distributed across multiple sources $\mathcal{S}$. In this case, the client and the intermediate nodes need to select properly the face over which they send the Interest for each segment, such that it reaches the right source. This is done using the information stored in the FIB table. However, keeping the FIB tables of all the nodes updated for each segment of $C_{p}$ does not scale well, in particular in large networks and in the presence of unreliable sources that can become available/unavailable at any moment.

\item \textit{Single-source multicast:} Let us now consider the case where a single source stores the $N$ segments that compose the content object $C_{p}$, but more than one client is interested in $C_{p}$, as illustrated in Fig.~\ref{fig:example_2}. In order to minimize the time needed for each client to receive the complete set of segments that compose $C_{p}$, while also minimizing the number of duplicated transmissions of the same segment in the network, the segments need to travel over cost-efficient multicast distribution trees~\cite{Wu2004}. In CCN, this means that each node of the network should know where each Interest $i_{p,n}$ should be forwarded such that all the Interests for the segment $c_{p,n}$ from different clients are aggregated in the optimal point in the network that reduces the number of copies of $c_{p,n}$ transmitted before reaching all the clients interested in it. However, finding the optimal set of multicast distribution trees in a distributed manner is a very complicated task~\cite{Wu2004}, which requires the knowledge of the network topology. In the simple example shown in Fig.~\ref{fig:example_2}, if all the clients send the Interest $i_{p,n}$ over the LTE face, the segment $c_{p,n}$ will be transmitted from the source to the LTE network and then to the clients. However, if some of the clients decide to send the Interest $i_{p,n}$ over the WiFi face, the segment $c_{p,n}$ will also be transmitted from the source to the WiFi network, wasting resources that could have been used to transmit another segment.

\item \textit{Multi-source multicast:} In this scenario, both of the above mentioned problems manifest, as it is a combination of both scenarios. Hence, it is more challenging for CCN to efficiently deliver the content to the clients.

\end{itemize}

\subsection{Enabling network coding in CCN}
\label{sec:nc_ccn}

The shortcomings of the CCN architecture discussed in Section \ref{sec:motivation} can be dealt with by enabling network coding~\cite{Ahlswede2000}, a technique in which the segments delivered to the clients are coded at sources and intermediate nodes. The key idea behind introducing network coding in CCN is that clients no longer need to request specific data segments, but rather encoded data segments as they all have the same amount of information. This removes the need to coordinate the forwarding of Interests and leads to a more efficient use of the available network bandwidth.

Differently from the original CCN where an Interest message $i_{p,n}$ requests a specific segment $c_{p,n}$, in a network coding enabled CCN variant an Interest message $\hat{i}_{p}$ requests a coded segment $\hat{c}_{p}$, without specifying a particular segment id. The sources or intermediate nodes can reply to these Interests with network coded segments, generated by combining the segments in the CS of the node that match the name prefix $p$. In matrix form, this can be expressed as $\hat{c}_{p} = \mathbf{A} \cdot \mathbf{CS_{p}}$, where $A$ is a vector of coding coefficients drawn from a Galois field, and $\mathbf{CS_{p}}$ is a vector of the segments in the CS of the node that match the name prefix $p$. When the coding coefficients in $\mathbf{A}$ used to combine the packets are chosen uniformly at random from a large enough Galois field, the generated segments have high probability of being linearly independent, and thus innovative. Whenever a client interested in a content object $C_{p}$ collects $N$ innovative coded segments $\hat{c}_{p}$, it can decode the original segments that compose $C_{p}$.

To illustrate the benefits that network coding brings to CCN, let us revisit the scenarios described in Section \ref{sec:motivation}.

\begin{itemize}

\item \textit{Multi-source unicast:} Differently to the original CCN, when network coding is allowed, the client and the intermediate nodes do not need to know over which face they can reach a particular source, since they send Interests for coded segments rather than for specific segments. This implies that the FIB of the clients and intermediate nodes do not need an entry for each segment, but only a single entry for the name prefix leading Interests towards the sources is enough. Each source then reply to these Interests with coded segments $\hat{c}_{p}$, generated by combining the segments that match the name prefix $p$.

\item \textit{Single-source multicast:} When network coding is allowed the clients do not need to coordinate what Interests they send over each face, since all of them are for coded segments. Thus, when all the clients send an Interest $\hat{i}_{p}$ requesting a coded segment over a face (\textit{e.g.}, LTE) it will be aggregated by the intermediate nodes, and only one Interest requesting a coded segment will reach the source.

\item \textit{Multi-source multicast:} In this scenario, when network coding is allowed, neither clients nor intermediate nodes need to coordinate the forwarding of the Interests, since they are for coded data and can be satisfied by any coded segment.

\end{itemize}

Network coding has also been shown to improve the throughput when bottlenecks are present in the network and the resiliency to packet erasures. In order to illustrate these benefits, let us consider the scenario in Fig.~\ref{fig:example_3}, commonly known as the butterfly network. We consider that two clients are interested in a content object $C_{p} = \{c_{p,1},c_{p,1}\}$ that consists of two segments. Each source holds only one of the segments. In this case, if network coding is not enabled, one of the clients will always receive the content with higher delay. This is because the link between nodes $r_{3}$ and $r_{4}$ becomes a bottleneck since the only way in which the clients $u_{1}$ and $u_{2}$ can get $c_{p,2}$ and $c_{p,1}$ respectively, is through node $r_{4}$. Thus, the Interests sent by the clients $u_{1}$ and $u_{2}$ cannot be aggregated in the node $r_{4}$, since they are for different segments. In contrast, when network coding is enabled, the Interests sent by the clients $u_{1}$ and $u_{2}$ can be aggregated in the node $r_{4}$, as they are both for coded data. If the node $v_{3}$ applies network coding to the segments received from the sources, the resulting coded segment will be useful for both clients $u_{1}$ and $u_{2}$.


\subsection{Challenges}
\label{sec:nc_ccn_consequences}

As discussed in Section \ref{sec:nc_ccn}, enabling network coding in CCN nodes brings benefits that can potentially improve the performance of content object retrieval under certain scenarios. However, some issues arise when the Interest messages do not specify the segment id.

One of the issues that arises is that any node that has a single coded segment $\hat{c}_{p}$ cached in its CS will reply with it to all the Interests $\hat{i}_{p}$, as the name prefix in the Interest matches that of the cached segment $\hat{c}_{p}$. This is undesirable, since the intermediate nodes will always reply with the same cached segment $\hat{c}_{p}$, while clients need to receive $N$ innovative coded segments in order to decode the original segments. Therefore, the intermediate nodes need a way to determine when they cannot provide a coded segment that is innovative to client, and thus a new coded segment need to be retrieved. In \cite{Montpetit2012} the authors propose to solve this problem by allowing the clients to include information about the coded segments they have collected so far. Intermediate nodes will reply to an Interest only if they can provide innovative information. However, it is not clear how intermediate nodes can aggregate Interests with different information from the clients.

Another challenge that emerges when network coding is enabled in CCN is related to the \textit{pipelining} procedure, \textit{i.e.}, a client sending multiple concurrent Interests for different segments of the same content object. In the original CCN when a node receives an Interest that it cannot satisfy with content stored in its CS, the node checks its PIT. If the node finds an entry in the PIT indicating that an Interest for the same name has been received previously over the same face, it will consider this new Interest as a duplicate and will not forward it. Since Interests for different segments will have different names, as the segment id is appended to the name prefix, pipelining is supported. However, when network coding is enabled in CCN, concurrent Interests for the same content object will be considered duplicated by the intermediate nodes, as Interests for different (linearly independent) coded segments of the same object will have the same name.

\section{The NetCodCCN Protocol}
\label{sec:netcodccn}

In this section, we present NetCodCCN, a practical implementation of a network coding enabled content-centric networking protocol. We build our proposal on the CCN~\cite{Jacobson2009} architecture. We start by defining the content segmentation and naming scheme in NetCodCCN. Then, we describe how Interests and Data messages are processed in our proposed protocol.

\subsection{Content Segmentation}
\label{sec:netcodccn_segmentation}

As in CCN, in NetCodCCN content objects are split into smaller segments, $C_{p} =\{ c_{p,1},\dots, c_{p,N} \}$, that fit into Data messages. Network coded segments $\hat{c}_{p}$ are generated by the sources by randomly combining the set of $L$ segments, $L \leq N$,  with name prefix $p$ that are stored in their CS. This set of segments is denoted as $\mathbf{CS_{p}}$. Thus, $\hat{c}_{p} = \sum_{l=0}^{l = L} a_{l} \cdot c_{p,l}$, where $a_{l}$ is a randomly selected coding coefficient and $c_{p,l}$ in sthe $l$th segment in $\mathbf{CS_{p}}$. As the segments $\mathbf{CS_{p}}$ used to create $\hat{c}_{p}$ and the coding coefficients $a_{l}$ are only known by the node performing network coding, information about both the segments and the coding coefficients should be included in each network coded segment, in order to allow decoding by the clients. In NetCodCCN, this is accomplished by following the approach presented in~\cite{Chou2007}, where an encoding vector $g_{n}$ is associated to the segment $c_{p,n} \in C_p$. The initial value of this encoding vector is an $n$th unit vector, meaning it is a vector that has value 1 in the $n$th position and 0 otherwise. Network coding operations are performed on both the segments and their corresponding encoding vectors, generating $\hat{c}_{p,g} = \sum_{l=0}^{l = L} a_{l} \cdot c_{p,l}$, where $\hat{c}_{p,g}$ is a coded vector associated with the encoding vector $g$ and $g = \sum_{l=0}^{l = L} a_{l} \cdot g_{l}$. Intermediate nodes generate network coded segments as the sources, but they combine segments that have been already coded, thus $\hat{c}_{p,g} = \sum_{l=0}^{l = L} a_{l} \cdot \hat{c}_{p,g_l}$, where $g = \sum_{l=0}^{l = L} a_{l} \cdot \hat{g}_{l}$ .

The clients and intermediate nodes keep track of the received innovative encoding vectors $g_{l}$, $\mathbf{G_{p}}=[g_{1};\dots;g_{L}]$, so that the original set of segments can be retrieved by performing Gaussian elimination when the matrix $\mathbf{G_{p}}$ is full rank, \emph{i.e.}, it contains $L=N$ linearly independent coding vectors.

The use of the encoding vectors introduces a communication overhead that leads to waste of network resources, especially when the content object is segmented in a large number of content segments. To limit this overhead, we adopt the concept of \emph{generations}~\cite{Chou2007}, where the original set of segments that compose $C_{p}$ is partitioned into smaller groups of segments, \emph{i.e.}, generations, and the coding operations are restricted only between segments that belong to the same generation. For example, let us consider that $C_{p}$ is partitioned into $K$ generations of $H_{k}$ segments each one. Hence, the $H_{k}$ segments of the $k$th generation are denoted as as $C_{p,k} =\{ c_{p,k,1}, ... , c_{p,k,H_{k}} \}$, where $k$ is the \emph{generation id}. The size of the generation, $H_{k}$, controls the tradeoff between the decoding delay, the packet diversity and the overhead required to communicate the encoding vector. Overall, the encoding vectors do not pose any limitations to our system as there are approaches to compress them efficiently \cite{Thomos2012,Lucani2014}. In order to avoid mixing packets from different generations, the segments are tagged with the generation id.

\subsection{Content Naming}
\label{sec:netcodccn_naming}

From the discussion above, it is obvious that the naming in NetCodCCN should have two additional components, \emph{i.e.}, the encoding vector $g$ and the generation id $k$. Let us consider a data object $C_{p}$ with content name $p =$ \texttt{/provider/videos/largevideo.h264}. This $C_{p}$ is partitioned into $K$ generations of $H_{k}$ segments each one. Thus, in NetCodCCN the first segment of the $k$th generation, $c_{p,k,1}$, associated with the unit vector $[1,0,0,0]$ when $H_{k}$ = 4, is named $\{p,k,1\}$ = \texttt{/provider/videos/largevideo.h264/k/1000}.


For the sake of clarity, and without loss of generality, hereafter we consider that the name prefix $p$ in $\hat{c}_{p,g}$, considers both the name prefix and the generation id. Note that, the proposed naming scheme is compatible with the original CCN and can support the delivery of non coded segments.

\subsection{Interest Message Processing}
\label{sec:netcodccn_intfwd}

Similarly to CCN, in our protocol the data communication is triggered by the clients who send Interest messages $\hat{i}_{p}$ for data with name prefix $p$. In the proposed NetCodCCN protocol, the Interests have a \textit{NetworkCodingAllowed} field that takes the value ``1'' when network coded segments are expected, otherwise, the field is not present or its value is set to ``0''. Nodes receiving an Interest with the \textit{NetworkCodingAllowed} field activated process the Interest messages following the NetCodCCN procedure explained below and summarized in Algorithm \ref{algo:interest_processing}. Otherwise, the Interests are treated following the original CCN procedures.

\begin{algorithm}
	\caption{Interest Processing in NetCodCCN}
	\label{algo:interest_processing}
	\begin{algorithmic}[1]
		\REQUIRE $\hat{i}_{p},f$, $\mathbf{CS_{p}} \gets$ segments that match $p$ in the CS
		\IF {Decoded($p$)}
				\STATE $\hat{c}_{p,g} \gets \sum_{l=0}^{l = L} a_{l} \cdot \hat{c}_{p,g_l}$
				\STATE Send segment $\hat{c}_{p,g}$ over face $f$
		\ELSE
				\STATE $\iota^{p,f} = Rank(\mathbf{G_p}) - n_{sent}^{p,f}$
				\IF {$\iota^{p,f} > 0$}
					\STATE $\hat{c}_{p,g} \gets \sum_{l=0}^{l = L} a_{l} \cdot \hat{c}_{p,g_l}$
					\STATE Send segment $\hat{c}_{p,g}$ over face $f$
				\ELSE
					\STATE InsertPIT ($p,f$)
					\IF {$n_{p}^{pending} \leq n_{p,f}^{pending}$}
						\STATE PropagateInterest($\hat{i}_{p}$)
					\ENDIF
				\ENDIF
		\ENDIF
	\end{algorithmic}
\end{algorithm}

When a node $v \in \mathcal{V}$ receives an Interest $\hat{i}_{p}$ for a network coded segment over a face $f$, it either \textit{(i)} replies to the Interest with a coded segment generated with the set of segments $\mathbf{CS_{p}}$; or \textit{(ii)} forwards the Interest to other nodes in order to receive a new linearly independent segment. This is further explained in the following.

\textbf{Replying an Interest:} The node $v$ replies to an Interest $\hat{i}_{p}$ when \textit{(i)} it has collected enough network coded segments to decode the original set of segments $C_{p}$; or when \textit{(ii)} a segment generated by node $v$ has high probability to be innovative for the node connected through face $f$ from where the Interest arrived. The number of coded segments $\iota^{p,f}$ that can be generated by the node $v$ and have high probability to be innovate for the node connected through face $f$, is given by $\iota^{p,f} = rank(\mathbf{G_p}) - \delta_{sent}^{p,f}$. The parameter $\delta_{sent}^{p,f}$ denotes the number of segments belonging to $\mathbf{CS_p}$ that have been previously sent over face $f$. When $\iota^{p,f}$ is higher that 0, $v$ generates a new coded segment and sends it over face $f$.

\begin{figure}
	\centering
		\includegraphics[width=0.30\textwidth]{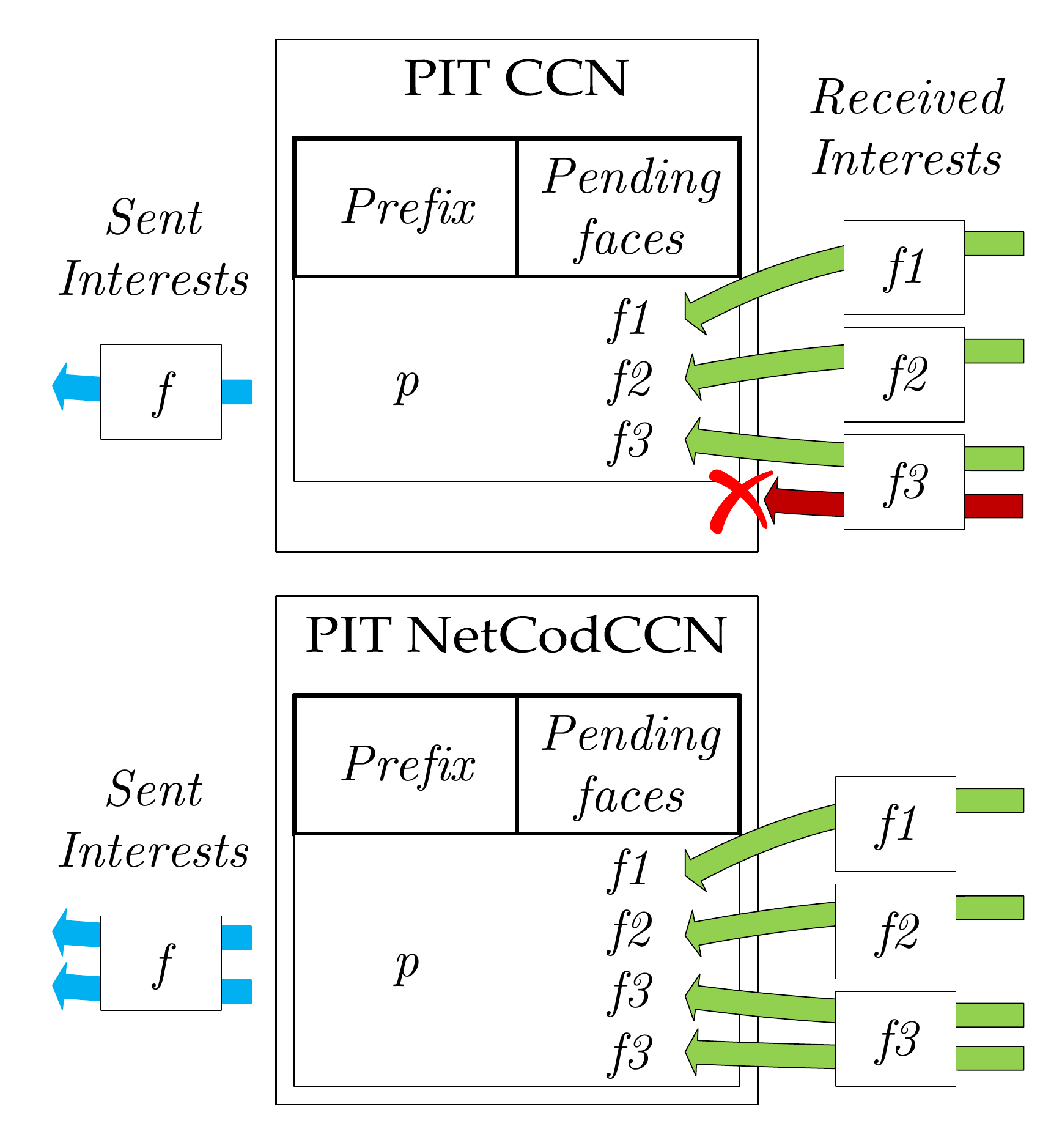}
	\caption{Comparison of the PIT in CCN and in NetCodCCN.}
	\label{fig:pit}
\end{figure}

\textbf{Forwarding an Interest:} The node $v$ forwards an Interest $\hat{i}_{p}$ to its neighbors when $\iota^{p,f}$ is equal to 0, since it needs a new segment that increases the rank of $\mathbf{G_p}$ before it can reply to Interest $\hat{i}_{p}$. As in CCN, prior to forwarding an Interest, the node $v$ checks its PIT. However, in order to support pipelining, in NetCodCCN the PIT verification procedure is modified. Specifically, if the node finds a matching PIT entry indicating that an Interest for the same name prefix $p$ has been previously received over the same face $f$, the Interest is not considered duplicated, but it is treated as a request for additional network coded segments from the same face, as shown in Fig. \ref{fig:pit}. This means that the face $f$ can appear multiple times in the PIT entry for the name prefix $p$. To decide whether the Interest $\hat{i}_{p}$ should be forwarded, the node $v$ computes the number $\nu_{p}$ of innovative coded segments matching the name prefix $p$ that it is expecting to receive before the Interest $\hat{i}_{p}$ expires. If $\nu_{p} > n_{p,f}^{pending}$, where $n_{p,f}^{pending}$ is the number of Interest received over face $f$ that are pending for a reply, the node $v$ does not forward the Interest $\hat{i}_{p}$, as it will receive enough coded segments to satisfy all the pending Interests, including the received Interest $\hat{i}_{p}$. Otherwise, if $\nu_{p} \leq n_{p,f}^{pending}$, the node $v$ forwards the Interest.

To compute the expected value of $\nu_{p}$, the node $v$ needs a probabilistic model that can take into consideration the loss rate of the system, the delays that segments may suffer, etc. For the sake of simplicity, we make the assumption that nodes follow a simple model in which any forwarded Interest brings an innovative segment before its expiration. In this case, $\nu_{p} = n_{p}^{pending}$, where $n_{p}^{pending}$ is the total number of pending interests for the name prefix $p$. Thus, the node will forward the Interest if $n_{p}^{pending} \leq n_{p,f}^{pending}$.

When a segment with name prefix $p$ is removed from the CS of node $v$ (\textit{e.g.}, when the CS eviction policy decides that segments of name prefix $p$ needs to be removed from the cache), the corresponding vector should also be removed from $\mathbf{G_p}$, and $\delta_{sent}^{p,f}$ needs to be decreased by 1 for all the faces.

\subsection{Data Message Processing}
\label{sec:netcodccn_objfwd}

The use of network coding in CCN imposes also modifications in the Data message forwarding procedure. Specifically, when a node $v \in \mathcal{V}$ receives a coded segment $\hat{c}_{p,g}$ on face $f$, it should determine whether $\hat{c}_{p,g}$ is innovative. The segment $\hat{c}_{p,g}$ is innovative for the node $v$ if the encoding vector $g$ is linearly independent in $\mathbf{G_{p}}$, \textit{i.e.}, if it increases the rank of $\mathbf{G_{p}}$. A non innovative segment is considered as a duplicated segment and therefore it is discarded by the node $v$. When the segment $\hat{c}_{p,g}$ is innovative, the node $v$ inserts it into its CS, and updates the encoding matrix $\mathbf{G_{p}}$ that should now contain the received encoded vector $g$. Then, the node $v$ checks its PIT. If the node $v$ finds a matching PIT entry, meaning that an Interest with name prefix $p$ is pending, it generates a network coded segment $\hat{c}_{p,g'} = \sum_{l=0}^{l = L} a_{l}\cdot\hat{c}_{p,g_l}$ and sends it once over each of the faces specified in the PIT entry. It is important to note that since the face $f$ can appear multiple times in the PIT entry for the name prefix $p$, as a consequence of allowing pipelining, a coded segment sent over face $f$ consumes only one of the appearances of $f$ in the corresponding PIT entry. If no matching PIT entry for the name prefix $p$ is found, the received coded segment is considered unsolicited and is not further transmitted. However, it can be kept in the CS, according to the CS insertion policy, as it can be useful to serve future Interests. This procedure is outlined in Algorithm \ref{algo:content_processing}.

\begin{algorithm}
	\caption{Content Processing in NetCodCCN}
	\label{algo:content_processing}
	\begin{algorithmic}[1]
		\REQUIRE $\hat{c}_{p,g}$
		\IF {$g$ increases the rank of $\mathbf{G_{p}}$}
			\STATE Insert $\hat{c}_{p,g}$ in the CS
			\STATE $PIT_{p} \gets$ PIT entry for prefix $p$
			\IF {$PIT_{p} = \varnothing$}
				\STATE Discard $\hat{c}_{p,g}$
			\ELSE
				\FORALL {$f$ in $PIT_{p}$}
					\IF {$f$ has not been served}
						\STATE $\hat{c}_{p,g'} = \sum_{l=0}^{l = L} a_{l}\cdot\hat{c}_{p,g_l}$
						\STATE Send segment $\hat{c}_{p,g'}$ over face $f$
						\STATE $n_{sent}^{p,f} \gets n_{sent}^{p,f} + 1$
						\STATE Remove one appareace of $f$ from $PIT_{p}$
					\ENDIF
				\ENDFOR
			\ENDIF
		\ELSE
			\STATE Discard $\hat{c}_{p,g}$
		\ENDIF
	\end{algorithmic}
\end{algorithm}

It is important to note that the application of network coding adds complexity to the Interest and Data message processing in NetCodCCN. Performing algebraic operations on the segments before forwarding them adds, effectively, some complexity to the CCN node. In particular, as we have seen in Section \ref{sec:netcodccn_segmentation}, a node generates a new coded segment $\hat{c}_{p,g'}$ as $\hat{c}_{p,g'} = \sum_{l=0}^{l = L} a_{l}\cdot\hat{c}_{p,g_l}$. If we consider that the operations are performed in a Galois field of size $2^8$ and that segments are of size $X$ symbols, each time a node needs to generate a new coded segment, it should perform $X \cdot L$ multiplications and $X \cdot (L - 1)$ additions. This complexity does not pose limitations to our scheme as there are efficient implementations of network coding~\cite{Pedersen2013}. Further, as it has been shown in~\cite{Zhang2011}, a network coding coder and decoder can operate at wire-speed with rates of up to 1000Mbps.

\section{NetCodCCN Evaluation}
\label{sec:systemevaluation}

\begin{figure*}[!t]
\centering
	%
	%
	\minipage{0.32\textwidth}
	\begin{tikzpicture}
		\tikzstyle{every node}=[font=\small]
		\begin{axis}[%
			width=0.7\textwidth,
			height=0.5\textwidth,
			at={(0\textwidth,0\textwidth)},
			scale only axis,
			xmin=5,
			xmax=10,
			xlabel={Bottleneck link capacity [Mbps]},
			xmajorgrids,
			ymin=0.95,
			ymax=2,
			ylabel={Normalized delay ($d$)},
			ymajorgrids,
			ytick={1,1.25,1.5,1.75,2},
			legend style={at={(0.97,0.85)},anchor=north east,legend cell align=left,align=left,draw=white!15!black}
			]
			\addplot [color=blue,solid,mark=diamond,mark options={solid}]
				table[row sep=crcr]{%
			5	1.00\\
			6.25	1.00\\
			7.5	1.00\\
			8.75	1.00\\
			10	1.0\\
			};
			\addlegendentry{NetCodCCN};

			\addplot [color=red,solid,mark=asterisk,mark options={solid},error bars/.cd,y dir=both,y explicit]
				table[row sep=crcr]{%
				5 1.20\\
				6.25	1.11\\
				7.5	1.01\\
				8.75	1.00\\
				10	1.00\\
				};
			\addlegendentry{CCNx-LS};

			\addplot [color=green,solid,mark=triangle,mark options={solid}]
				table[row sep=crcr]{%
			5	1.91\\
			6.25	1.91\\
			7.5	1.91\\
			8.75	1.91\\
			10	1.91\\
			};
			\addlegendentry{CCNx-PS};

			\addplot [color=orange,solid,mark=square,mark options={solid}]
				table[row sep=crcr]{%
			5	1.90\\
			6.25	1.92\\
			7.5	1.92\\
			8.75	1.90\\
			10	1.92\\
			};
			\addlegendentry{CCNx-DS};
		\end{axis}
	\end{tikzpicture}%
	\caption{Normalized delivery delay versus the capacity of the bottleneck link in the butterfly network.}
	\label{fig:dd_capacity_butterfly}
	\endminipage\hfill
	%
	%
	\minipage{0.32\textwidth}
	\begin{tikzpicture}
		\tikzstyle{every node}=[font=\small]
		\begin{axis}[%
			width=0.7\textwidth,
			height=0.5\textwidth,
			at={(0\textwidth,0\textwidth)},
			scale only axis,
			xmin=2,
			xmax=30,
			xlabel={Pipeline size},
			xmajorgrids,
			ymin=0.95,
			ymax=2,
			ylabel={Normalized delay ($d$)},
			ymajorgrids,
			ytick={1,1.25,1.5,1.75,2},
			xtick={2,5,10,15,20,25,30},
			legend style={at={(0.03,0.97)},anchor=north west,legend cell align=left,align=left,draw=white!15!black}
			]
			\addplot [color=blue,solid,mark=diamond,mark options={solid}]
				table[row sep=crcr]{%
			2	1.00\\
			3	1.00\\
			4	1.00\\
			5	1.00\\
			10	1.00\\
			15	1.00\\
			20	1.00\\
			25	1.00\\
			30	1.00\\
			};
			\addlegendentry{NetCodCCN};

			\addplot [color=red,solid,mark=asterisk,mark options={solid}]
				table[row sep=crcr]{%
			2	1.71\\
			3	1.43\\
			4	1.24\\
			5	1.19\\
			10	1.18\\
			15	1.37\\
			20	1.54\\
			25	1.81\\
			30	1.95\\
			};
			\addlegendentry{CCNx-LS};
		\end{axis}
	\end{tikzpicture}%
	\caption{Normalized delivery delay versus the pipeline size in the butterfly network.}
	\label{fig:dd_pipeline_butterfly}
	\endminipage\hfill
	%
	%
	\minipage{0.32\textwidth}
	\begin{tikzpicture}
	\tikzstyle{every node}=[font=\small]
		\begin{axis}[%
			width=0.7\textwidth,
			height=0.5\textwidth,
			at={(0\textwidth,0\textwidth)},
			scale only axis,
			xmin=0,
			xmax=30,
			xlabel={Error rate [\%]},
			xmajorgrids,
			ymin=0.95,
			ymax=3,
			ylabel={Normalized delay ($d$)},
			ymajorgrids,
			ytick={1,1.5,2,2.5,3},
			legend style={at={(0.03,0.97)},anchor=north west,legend cell align=left,align=left,draw=white!15!black}
			]
			\addplot [color=blue,solid,mark=diamond,mark options={solid}]
				table[row sep=crcr]{%
			0	1.00\\
			5	1.04\\
			10	1.17\\
			15	1.38\\
			20	1.42\\
			25	1.49\\
			30	1.69\\
			};
			\addlegendentry{NetCodCCN};

			\addplot [color=red,solid,mark=asterisk,mark options={solid}]
				table[row sep=crcr]{%
			0	1.32\\
			5	1.41\\
			10	1.57\\
			15	2.05\\
			20	2.18\\
			25	2.38\\
			30	2.81\\
			};
			\addlegendentry{CCNx-LS};
		\end{axis}
	\end{tikzpicture}%
	\caption{Normalized delivery delay versus the error rate in the butterfly network.}
	\label{fig:dd_error_butterfly}
	\endminipage\hfill
\end{figure*}

In this section, we evaluate the performance of NetCodCCN in various scenarios, and compare the results to the performance of the standard CCN. First, we describe the simulation setup. Then, we evaluate the performance of NetCodCCN in the butterfly network. This toy network provides a controllable environment which permits to verify the expected behavior of NetCodCCN, and facilitates illustration of its benefits. Finally, we present the simulation results in a more realistic network topology, which is generated based on real network measurements taken from the Planetlab project \cite{Planetlab2015}.

\subsection{Simulation Setup}
We implemented NetCodCCN by integrating the cahnges to the CCN architecture described in Section \ref{sec:netcodccn} into the CCNx 0.8.2\cite{Ccnx2015} code, and we compare its performance to that of the unmodified CCNx. The network topology is simulated using the NS-3 network simulator \cite{ns3}. The software forwarders/routers for CCNx and NetCodCCN are installed on NS-3 nodes using the Direct Code Execution framework (DCE) \cite{ns3DCE}.

We consider that the clients are interested in a content object composed of $N=100$ segments. The size of each segment is 5KB. The data segments are stored in a set of sources that are connected to the clients through a network of intermediate nodes. We consider that the intermediate nodes have sufficient CS space to store all the incoming data. We assume that the $N= 100$ source data segments comprise a single generation, \emph{i.e.}, $H = N$ and $K=1$. The finite field in which the network coding operations are performed is of size $2^8$. In order to evaluate our protocol in a challenging scenario, we consider that all the clients send Interests for the content segments during the same interval of time. In this way, we demonstrate that by using our protocol, nodes are able to aggregate Interests adequately.

For the evaluation of CCNx, we consider the three main Interest forwarding strategies implemented in CCNx 0.8.2, and described in~\cite{Ccnx2015-ccndc}:
\begin{itemize}
	\item The \textit{default} (DS) strategy selects the fastest responding face based on the face statistics.
	\item The \textit{loadsharing} (LS) strategy distributes the Interest forwarding load over all the available faces, sending each Interest over the face with the smallest pending Interest queue.
	\item The \textit{parallel} (PS) strategy sends the Interests in parallel over all the faces indicated on the FIB.
\end{itemize}
For the evaluation of NetCodCCN, we always consider the \textit{parallel} strategy, since by sending a single Interest over all its faces the client can receive multiple useful segments, \emph{i.e.}, linearly independent.

To evaluate the performance of NetCodCCN, we measure the time $\Delta t_{measured}$ that a client needs in order to get the $N$ segments available at the sources. In the standard CCN, $\Delta t_{measured}$ is defined as the elapsed time between the transmission of the first Interest and the reception of the $N$th missing segment. In NetCodNCC, $\Delta t_{measured}$ is defined as the elapsed time between the transmission of the first Interest and the reception of the $N$th linearly independent network coded segment, which permits to decode the whole generation of segments. We consider that clients can have heterogeneous network resources, thus, in order to make a fair comparison of the delivery delay, we define the \emph{normalized delivery delay} as $d = \Delta t_{measured} / \Delta t_{min}$, where $\Delta t_{min}$ is the theoretical lower bound on the time that a client would need in order to receive all the segments if it was alone in the network and was able to receive at max-flow rate. Thus, a normalized delivery delay equal to 1 means that the client was able to receive the complete set of packets at the maximum rate. Note that $\Delta t_{measured} \geq \Delta t_{min} $, or equivalently, $d \geq 1$ always holds.

\begin{figure*}[!t]
\centering
	%
	%
	\minipage{0.32\textwidth}
	\begin{tikzpicture}
	\tikzstyle{every node}=[font=\small]
		\begin{axis}[%
			width=0.7\textwidth,
			height=0.5\textwidth,
			at={(0\textwidth,0\textwidth)},
			scale only axis,
			xmin=0,
			xmax=100,
			xlabel={Duplication probability $\phi$ [\%]},
			xmajorgrids,
			ymin=0.95,
			ymax=4,
			ylabel={Normalized delay ($d$)},
			ymajorgrids,
			ytick={1,1.5,2,2.5,3,3.5,4},
			legend style={at={(0.97,0.97)},anchor=north east,legend cell align=left,align=left,draw=white!15!black}
			]
			\addplot [color=blue,solid,mark=diamond,mark options={solid}]
				table[row sep=crcr]{%
			0	1.04	\\
			10	1.00	\\
			20	1.00	\\
			30	1.00	\\
			40	1.00	\\
			50	1.00	\\
			60	1.00	\\
			70	1.00	\\
			80	1.00	\\
			90	1.00	\\
			100	1.00	\\
			};
			\addlegendentry{NetCodCCN};

			\addplot [color=red,solid,mark=asterisk,mark options={solid}]
				table[row sep=crcr]{%
			0	3.39	\\
			10	3.24	\\
			20	3.11	\\
			30	2.50	\\
			40	2.32	\\
			50	2.02	\\
			60	1.33	\\
			70	1.33	\\
			80	1.33	\\
			90	1.33	\\
			100	1.33	\\
			};
			\addlegendentry{CCNx-LS};

			\addplot [color=green,solid,mark=triangle,mark options={solid}]
				table[row sep=crcr]{%
			0	2.00	\\
			10	2.04	\\
			20	2.03	\\
			30	1.93	\\
			40	2.02	\\
			50	2.02	\\
			60	1.91	\\
			70	1.91	\\
			80	1.91	\\
			90	1.91	\\
			100	1.91	\\
			};
			\addlegendentry{CCNx-PS};
		\end{axis}
	\end{tikzpicture}%
	\caption{Normalized delivery delay versus the source content duplication probability in the butterfly network.}
	\label{fig:dd_duplication_butterfly}
	\endminipage\hfill
	%
	%
	\minipage{0.32\textwidth}
	\begin{tikzpicture}
		\tikzstyle{every node}=[font=\small]
		\begin{axis}[%
			width=0.7\textwidth,
			height=0.5\textwidth,
			at={(0\textwidth,0\textwidth)},
			scale only axis,
			xmin=1,
			xmax=5,
			xlabel={Number of clients},
			xmajorgrids,
			ymin=0.95,
			ymax=2.5,
			ylabel={Normalized delay ($d$)},
			ymajorgrids,
			legend style={at={(0.03,0.97)},anchor=north west,legend cell align=left,align=left,draw=white!15!black}
			]
			\addplot [color=blue,solid,mark=diamond,mark options={solid}]
				table[row sep=crcr]{%
			1	1.09\\
			2	1.11\\
			3	1.17\\
			4	1.12\\
			5	1.12\\
			};
			\addlegendentry{NetCodCCN};
			\addplot [color=red,solid,mark=asterisk,mark options={solid}]
				table[row sep=crcr]{%
			1	1.16\\
			2	1.47\\
			3	1.71\\
			4	1.9\\
			5	2.15\\
			};
			\addlegendentry{CCNx-LS};
			\addplot [color=orange,solid,mark=square,mark options={solid}]
				table[row sep=crcr]{%
			1	1.05\\
			2	1.34\\
			3	1.65\\
			4	1.79\\
			5	1.91\\
			};
			\addlegendentry{CCNx-DS};
		\end{axis}
	\end{tikzpicture}%
	\caption{Normalized delivery delay versus the number of clients in the network in the PlanetLab topology.}
	\label{fig:dd_nclients_planetlab}
	\endminipage\hfill
	%
	%
	\minipage{0.32\textwidth}
	\begin{tikzpicture}
		\tikzstyle{every node}=[font=\small]
		\begin{axis}[%
			width=0.7\textwidth,
			height=0.5\textwidth,
			at={(0\textwidth,0\textwidth)},
			scale only axis,
			xmin=0,
			xmax=5,
			xlabel={Error Rate [\%]},
			xmajorgrids,
			ymin=1,
			ymax=3.5,
			ylabel={Normalized delay ($d$)},
			ymajorgrids,
			ytick={1,1.5,2,2.5,3,3.5},
			legend style={at={(0.03,0.97)},anchor=north west,legend cell align=left,align=left,draw=white!15!black}
			]
			\addplot [color=blue,solid,mark=diamond,mark options={solid}]
				table[row sep=crcr]{%
			0	1.05\\
			1	1.08\\
			2	1.09\\
			3	1.11\\
			4	1.11\\
			5	1.13\\
			};
			\addlegendentry{NetCodCCN};
			\addplot [color=red,solid,mark=asterisk,mark options={solid}]
				table[row sep=crcr]{%
			0	1.16\\
			1	2.37\\
			2	2.43\\
			3	2.74\\
			4	2.77\\
			5	3.27\\
			};
			\addlegendentry{CCNx-LS};
		\end{axis}
	\end{tikzpicture}%
	\caption{Normalized delivery delay versus the segment transmission error rate in the PlanetLab topology.}
	\label{fig:dd_error_planetlab}
	\endminipage\hfill
\end{figure*}

\subsection{Butterfly Topology}

We begin by evaluating NetCodCCN in the butterfly topology presented in Fig.~\ref{fig:example_3}. We consider that every data segment is stored randomly in at least one of the two sources, and a copy of the same segment is also placed in the Content Store of the other source with a duplication probability $\phi \in [0,1]$. We set the capacity of every link in the network to $5Mbps$.

In the first set of experiments, we consider that $\phi = 1$, \emph{i.e.}, both sources hold a copy of each data segment in their CS. This corresponds to the single source multicast case presented in Section \ref{sec:motivation}. In this case, clients $u_{1}$ and $u_{2}$ can reach a copy of any segment over any of their faces. However, as explained in Section~\ref{sec:nc_ccn}, with the original CCN protocol the maximum performance can be achieved only if both clients coordinate and send Interest messages for the same segments over the faces that connect them to the node $r_{4}$. In contrast, when network coding is employed, the need for coordination is eliminated, since clients do not send Interests for a specific segment but rather for any network coded segment.

Fig.~\ref{fig:dd_capacity_butterfly} depicts the normalized delivery delay as a function of the capacity of the bottleneck link between nodes $r_{3}$ and $r_{4}$. We can see that NetCodCCN achieves the optimal performance in the whole range of link capacity values. This is due to the fact that network coding removes the need for coordinating the forwarding of Interest messages. In contrast, the CCN forwarding strategies perform poorly and only the LS strategy can achieve the performance of NetCodCCN but requires significantly higher link capacity. When the bottleneck link has the same capacity as all the other links, the average delivery time $d$ of CCNx-LS  is around 1.2 times the minimum delivery delay, $\Delta  t_{min}$. This is caused by the randomness introduced by the LS strategy when choosing the faces over which Interests are transmitted when all the faces have the same load. This creates two extreme cases. In one case, all the Interests sent by both clients to node $r_4$ are the same, thus $d$ tends to one. In the other case, all the Interests are different, thus $d$ tends to 1.33. This happens because each node will receive 2/3 of the segments through the link connecting them to the sources, and 1/3 over the face connecting them to the node $r_4$, which means that 2/3 of the total segments will travel on the bottleneck link. With the DS and the PS strategies, the average delivery time is close to 2, as expected. With the DS strategy, the face connecting the clients to the sources will be chosen as the best, and thus most of the segments will be received over that face. With the PS strategy, each client will forward every Interest over both faces, thus bringing one copy of each segment over each face.

We now investigate how the number of concurrent Interests that a client can send, also known as the pipeline size, affects the performance in terms of the average normalized delivery delay. As shown in Fig.~\ref{fig:dd_pipeline_butterfly}, the performance of CCN is optimized for a pipeline size value between 5 and 10, where the normalized delivery delay seen by the clients is 1.2. This is due to the fact that clients need to send at least 4 Interest messages over the faces connecting them to the node $r_{4}$ in order to create a continuous flow of segments. Since the LS strategy distributes the Interests over all available faces, a client has to send 4 Interests over each face while it also has sent 3 or 4 Interests over the other face, which amounts to 7 or 8 Interests in total. For smaller pipeline sizes, the continuous flow is not set, while for larger pipeline sizes the number of Interests sent over the bottleneck link increases, thus worsening the client coordination problem. In contrast, the performance of NetCodCCN is not affected by the pipeline size, as can be verified in Fig.~\ref{fig:dd_pipeline_butterfly}. This can be explained by the fact that NetCodCCN eliminates the necessity that the clients request the same segments over the bottleneck link. For the rest of the experiments, we choose a pipeline size of 10.

In Fig.~\ref{fig:dd_error_butterfly}, we depict the influence of the data segment loss rate on the performance of NetCodCCN and of the original CCN. We consider losses that are caused both by the transmission losses and the errors during the processing of the segments. We can see that the performance of CCN with the LS strategy degrades faster than the performance of NetCodCCN as the segment error rate increases. This is caused by the fact that in CCN, the client will be able to react to a segment loss only when the corresponding Interest expires, since any earlier re-transmission of an Interest with the same prefix will be prevented by the PIT. Instead, with NetCodCCN, the clients can send Interests for new coded segments until they have a sufficient number of coded segments in order to recover the original ones. It is important to note that the maximum amount of concurrent Interests that a client can send is controlled by the pipeline size.

Finally, we evaluate the performance of NetCodCCN for different values of the duplication probability $\phi$. This corresponds to the multi-source multicast case presented in Section~\ref{sec:motivation}. In CCNx, when $\phi < 1$, the clients should not only coordinate the requests sent over the bottleneck link as in the previous scenario, but they also should have the knowledge of the segments that each source stores, in order to avoid sending Interests over the face connecting them directly to the source that does not hold a copy of the requested segment. In Fig.~\ref{fig:dd_duplication_butterfly} we can see that CCN with the LS strategy takes 3.4 times longer to deliver all the segments to the clients, when each segment is stored only in one of the sources. When the PS strategy is employed, the clients do not need to know how the content is distributed since each Interest message is sent over both of its faces. However, since a copy of every segment will cross the bottleneck link, the traffic over the bottleneck link will be doubled compared to the network coding case. When the probability that the segments are stored in both sources increases, the performance of CCNx with the LS strategy improves, but eventually saturates at 1.2 times the minimum delay, which is consistent with the results depicted in Fig.~\ref{fig:dd_pipeline_butterfly}.

\subsection{Planetlab Topologies}

\begin{figure}[t]
	\centering
		\includegraphics[width=0.3\textwidth]{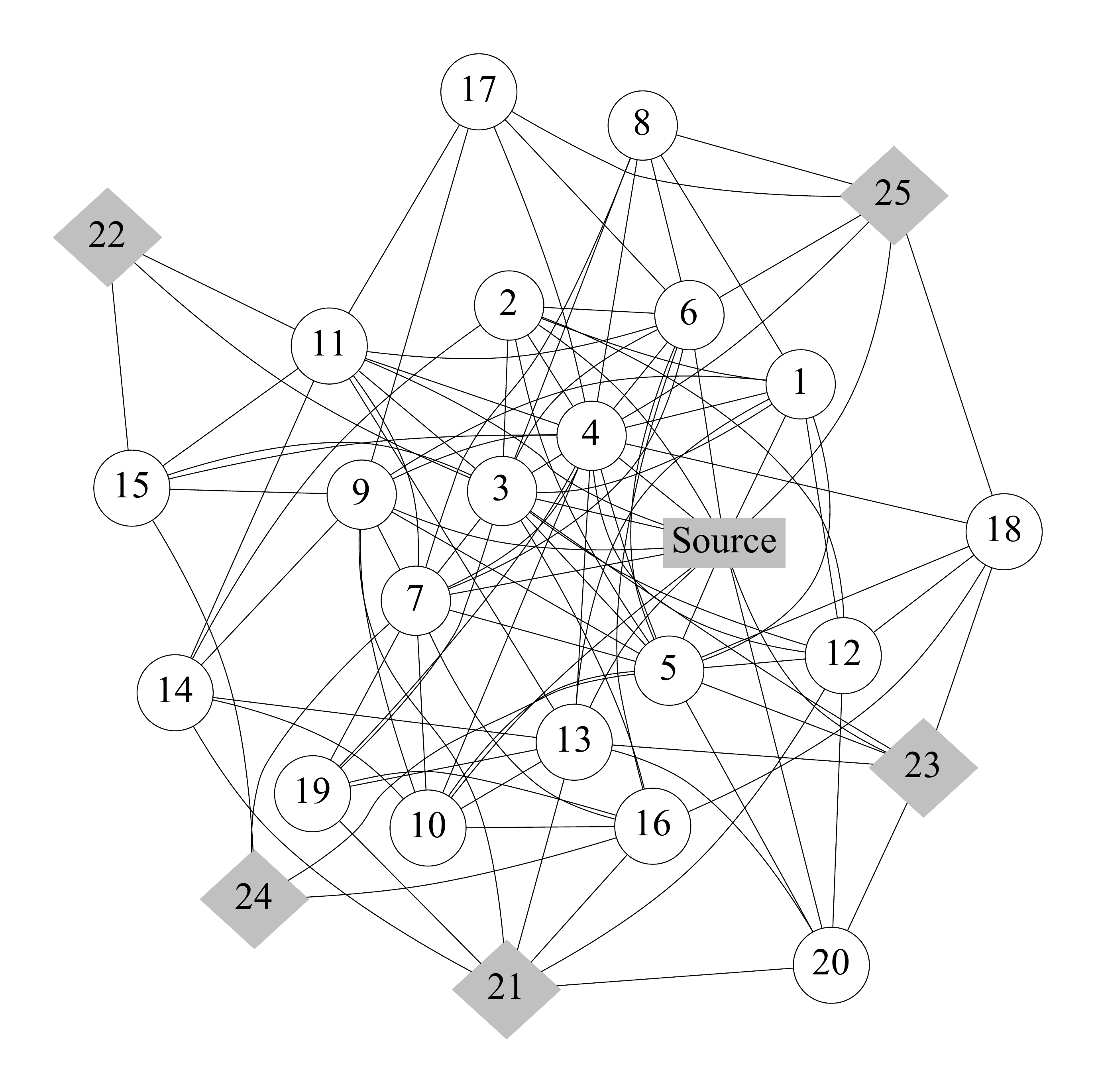}
	\caption{Planetlab topology used.}
	\label{fig:planetlab_topology}
\end{figure}

We now evaluate our protocol in more realistic network topologies captured by the PlanetLab project \cite{Planetlab2015}. We use the network topology shown in Fig.~\ref{fig:planetlab_topology} that consists of one source node, 5 client nodes and 20 intermediate nodes. The links connecting the nodes have a capacity of 12Mbps. The topology was generated using the procedure described in \cite{Cleju2011}. We measure the normalized delivery delay $d$ for each client and then compare its average.

First, we investigate how the performance is affected by the number of clients in the network. In Fig.~\ref{fig:dd_nclients_planetlab} we can see that with a single client (in this case node 24), NetCodCCN and CCNx perform similarly. In this case, network coding does not introduce any gains since there is only one client in the network and no losses are considered. However, the performance of CCNx starts to degrade with the introduction of more clients, as they start to compete for the network resources. In contrast, we can see that the performance of NetCodCCN does not deteriorate with the addition of new clients to the network topology. These results show that the NetCodCCN protocol uses more efficiently the available network resources.

We also evaluate how the error in segment transmission affects the performance of the NetCodCCN for larger topologies. For this evaluation, we choose to keep only one client, in order to compare the results with the performance of the CCN. As with the butterfly topology, we consider losses that are caused both by the transmission losses and the errors during the processing of the segments. In Fig.~\ref{fig:dd_error_planetlab} we can see that NetCodCCN maintains the delivery delay close to the expected one, while the performance of CCN degrades very fast with the introduction of errors. As in the butterfly topology, this fast degradation is due to the fact that when a segment is lost, the client needs to wait until the corresponding Interest expires before it can re-send a new one.

\section{Conclusions}
\label{sec:conclusions}
In this paper, we have presented NetCodCCN, a protocol that integrates network coding in CCN. Specifically, we have defined novel algorithms to process Interest and Data messages so that consider network coded content. In NetCodCCN, the clients express Interest messages for coded segments of a given prefix instead of asking a specific segment as in CCN. The network nodes combine the Data messages by means of RLNC before forwarding them in order to take advantage of the network diversity. Our protocol is able to \textit{(i)} simplify the aggregation of Interest for coded content; \textit{(ii)} reduce the number of duplicate segments; and \textit{(iii)} allow clients to send multiple Interests for the same content in parallel. The overall system has been tested in networks with multiple clients and sources, where we have observed large performance gains in terms of the time needed to retrieve the demanded content.

Our future research includes the investigation of optimal Interest forwarding strategies that enable flow control in the case of multiple different content objects. We will also consider the transmission of video content characterized by strict delivery deadlines. Furthermore, we will focus on enabling content security on NetCodCCN.

\section*{Acknowledgments}
This work has been partially funded by the  Swiss  National Science Foundation under grant number 149225.

\bibliographystyle{ieeetran}
\bibliography{infocom16}

\begin{thebibliography}{10}
\providecommand{\url}[1]{#1}
\csname url@samestyle\endcsname
\providecommand{\newblock}{\relax}
\providecommand{\bibinfo}[2]{#2}
\providecommand{\BIBentrySTDinterwordspacing}{\spaceskip=0pt\relax}
\providecommand{\BIBentryALTinterwordstretchfactor}{4}
\providecommand{\BIBentryALTinterwordspacing}{\spaceskip=\fontdimen2\font plus
\BIBentryALTinterwordstretchfactor\fontdimen3\font minus
  \fontdimen4\font\relax}
\providecommand{\BIBforeignlanguage}[2]{{%
\expandafter\ifx\csname l@#1\endcsname\relax
\typeout{** WARNING: IEEEtran.bst: No hyphenation pattern has been}%
\typeout{** loaded for the language `#1'. Using the pattern for}%
\typeout{** the default language instead.}%
\else
\language=\csname l@#1\endcsname
\fi
#2}}
\providecommand{\BIBdecl}{\relax}
\BIBdecl

\bibitem{Jacobson2009}
\BIBentryALTinterwordspacing
V.~Jacobson, D.~K. Smetters, J.~D. Thornton, M.~F. Plass, N.~H. Briggs, and
  R.~L. Braynard, ``{Networking Named Conten}t,'' in \emph{Proc. of ACM
  CoNEXT}, New York, NY, USA, 2009, pp. 1--12. [Online]. Available:
  \url{http://doi.acm.org/10.1145/1658939.1658941}
\BIBentrySTDinterwordspacing

\bibitem{Wu2004}
Y.~Wu, P.~Chou, and K.~Jain, ``{A Comparison of Network Coding and Tree
  Packing},'' in \emph{Proc. of IEEE ISIT'04}, June 2004, pp. 143--.

\bibitem{Ahlswede2000}
R.~Ahlswede, N.~Cai, S.-Y. Li, and R.~Yeung, ``{Network Information Flow},''
  \emph{IEEE Trans. Information Theory}, vol.~46, no.~4, pp. 1204--1216, Jul
  2000.

\bibitem{Montpetit2012}
\BIBentryALTinterwordspacing
M.-J. Montpetit, C.~Westphal, and D.~Trossen, ``{Network Coding Meets
  Information-Centric Networking: An Architectural Case for Information
  Dispersion Through Native Network Coding},'' in \emph{Proc. of the 1st ACM
  NoM Workshop}.\hskip 1em plus 0.5em minus 0.4em\relax New York, NY, USA: ACM,
  2012, pp. 31--36. [Online]. Available:
  \url{http://doi.acm.org/10.1145/2248361.2248370}
\BIBentrySTDinterwordspacing

\bibitem{Sundararajan2011}
J.~Sundararajan, D.~Shah, M.~Medard, S.~Jakubczak, M.~Mitzenmacher, and
  J.~Barros, ``Network coding meets tcp: Theory and implementation,''
  \emph{Proceedings of the IEEE}, vol.~99, no.~3, pp. 490--512, March 2011.

\bibitem{Wu2013}
\BIBentryALTinterwordspacing
Q.~Wu, Z.~Li, and G.~Xie, ``{Coding{C}ache: Multipath-Aware CCN Cache with
  Network Coding},'' in \emph{Proc. of the 3rd ACM ICN Workshop}.\hskip 1em
  plus 0.5em minus 0.4em\relax New York, NY, USA: ACM, 2013, pp. 41--42.
  [Online]. Available: \url{http://doi.acm.org/10.1145/2491224.2491240}
\BIBentrySTDinterwordspacing

\bibitem{Llorca2013}
J.~Llorca, A.~Tulino, K.~Guan, and D.~Kilper, ``{Network-Coded Caching-Aided
  Multicast for Efficient Content Delivery},'' in \emph{Proc. of IEEE ICC'13},
  June 2013, pp. 3557--3562.

\bibitem{Ccnx2015}
``{P}roject {CCN}x \textregistered, version 0.8.2,''
  http://www.ccnx.org/releases/ccnx-0.8.2/doc/.

\bibitem{Chou2007}
P.~Chou and Y.~Wu, ``{Network Coding for the Internet and Wireless Networks},''
  \emph{IEEE Signal Processing Magazine}, vol.~24, no.~5, pp. 77--85, Sept
  2007.

\bibitem{Thomos2012}
N.~Thomos and P.~Frossard, ``{Toward one Symbol Network Coding Vectors},''
  \emph{{IEEE Communications letters}}, vol.~16, no.~11, pp. 1860--1863, Nov.
  2012.

\bibitem{Lucani2014}
D.~E. Lucani, M.~V. Pedersen, J.~Heide, and F.~H.~P. Fitzek, ``{Fulcrum Network
  Codes: A Code for Fluid Allocation of Complexity},'' \emph{{available at
  http://arxiv.org/abs/1404.6620}}, 2014.

\bibitem{Pedersen2013}
M.~Pedersen, J.~Heide, P.~Vingelmann, and F.~Fitzek, ``Network coding over the
  $2^{32}-5$ prime field,'' in \emph{Communications (ICC), 2013 IEEE
  International Conference on}, June 2013, pp. 2922--2927.

\bibitem{Zhang2011}
M.~Zhang, H.~Li, F.~Chen, H.~Hou, H.~An, W.~Wang, and J.~Huang, ``A general
  co/decoder of network coding in hdl,'' in \emph{Network Coding (NetCod), 2011
  International Symposium on}, July 2011, pp. 1--5.

\bibitem{Planetlab2015}
``{P}lanet{L}ab,'' https://www.planet-lab.org/.

\bibitem{ns3}
``The network simulator - ns3,'' http://www.nsnam.org/.

\bibitem{ns3DCE}
``{D}irect {C}ode {E}xecution ({DCE}),''
  https://www.nsnam.org/overview/projects/\\direct-code-execution/.

\bibitem{Ccnx2015-ccndc}
``{C}{C}{N}{D}{C}(1) {M}anual {P}age, {P}roject {CCN}x \textregistered, version
  0.8.2,'' https://www.ccnx.org/releases/ccnx-0.8.2/doc/manpages/ccndc.1.html.

\bibitem{Cleju2011}
N.~Cleju, N.~Thomos, and P.~Frossard, ``{Selection of Network Coding Nodes for
  Minimal Playback Delay in Streaming Overlays},'' \emph{IEEE Trans.
  Multimedia}, vol.~13, no.~5, pp. 1103--1115, Oct 2011.

\end{thebibliography}

\end{document}